\documentclass[11pt,a4paper]{article}
\usepackage[T1]{fontenc}
\usepackage[utf8]{inputenc}
\usepackage{mathptmx}
\usepackage[scaled=0.92]{helvet}
\usepackage{courier}
\usepackage[margin=1in]{geometry}
\usepackage{graphicx}
\usepackage{booktabs}
\usepackage{caption}
\usepackage[expansion=false]{microtype}
\usepackage{setspace}
\usepackage[hidelinks,breaklinks]{hyperref}
\usepackage{url}
\title{\bfseries Computational Prosopography across a Millennium:\\
Mathematically Oriented Lineages Traced from the Fields Medalists}

\author{
Hiroyuki Chuma\thanks{Corresponding author: \texttt{chuma@iir.hit-u.ac.jp}. Institute of Innovation
Research, Hitotsubashi University, 2-1 Naka, Kunitachi, Tokyo 186-8603, Japan.}\\
\small Professor Emeritus, Hitotsubashi University and Seijo University
\and
Kanji Otsuka\\
\small Professor Emeritus, Meisei University
\and
Yoichi Sato\\
\small Shuhari System
}
\date{\today}

\begin{document}
\maketitle
\begin{abstract}\noindent

We reconstruct the mentor--student network through which documented scholarly training passed across roughly nine centuries, and subject both the network and the means of reconstructing it to source criticism. From Wikidata, which aggregates the Mathematics Genealogy Project and the MacTutor Archive, we extract approximately 470,000 mentor--student assertions, yielding a directed acyclic graph of 372,853 persons. Using all 64 historical Fields Medalists as a fixed, ex ante tracer set, backward traversal enumerates some 25.5 million distinct paths reaching 57 generations.

Three structural observations follow. Genealogical traffic through Leibniz forms an hourglass: thin upstream, 5.3 paths per node on average, and thick downstream, 53.4, a ratio near 10:1, with no counterpart at Newton, who lies on only four of the 64 lineages. Across a window centered on Leibniz, seven independently extracted predicate dimensions reorganize together, and recorded learned-society membership rises from 6.5 to 82.1 percent of the cohort. Upstream, 54 of the 64 lineages converge on the same five twelfth- and thirteenth-century Islamic and Byzantine scholars before terminating at an eleventh-century boundary we name the Monastery Wall.

We argue that such observations cannot be assessed without tool criticism. The traversal engine is algebraically reversible, so every ranking decision it makes can be reconstructed afterward. We characterize its measurement bias in closed form, show that the macro-structures survive switching that bias off, and report the family of lineages the traversal returns at different resolutions rather than a single ranked list.

\end{abstract}

\section{Introduction}

Historical network research has long insisted that a network is not found but made: the edges an analyst draws are artifacts of what a source recorded, why it recorded it, and what it was structurally incapable of recording.\footnote{Claire Lemercier, ``Formal Network Methods in History: Why and How?,'' in Social Networks, Political Institutions, and Rural Societies, ed. Georg Fertig (Turnhout: Brepols, 2015), 281--310; Marten Düring and Linda von Keyserlingk, ``Netzwerkanalyse in den Geschichtswissenschaften,'' in Prozesse: Formen, Dynamiken, Erklärungen, ed. Rainer Schützeichel and Stefan Jordan (Wiesbaden: Springer VS, 2015), 337--350.} This paper takes that insistence as its organizing problem rather than as a caveat appended to results.

The object we reconstruct is the mentor--student network of documented scholarly training, traced backward from the present to the edge of the medieval record. Academic genealogy, the chain ``A was taught by B, who was taught by C,'' has an unusual property among historical relations. It is transitive in a way that letters, citations, and co-membership are not, and it therefore composes into paths of great length. A single documented tie is a small fact. Twenty of them in sequence connect a living mathematician to a fourteenth-century schoolman. What that composition means, and how much of it is an artifact of the composing, is the question this paper is about.

The empirical scale is large enough to make the question nontrivial: from roughly 470,000 mentor--student assertions in Wikidata we enumerate some 25.5 million distinct paths across 57 generations. It is also large enough to defeat unaided intuition and the standard toolkit alike, which is why the paper spends as much space on the traversal as on the traversed. That emphasis is deliberate, and, we would argue, overdue in a field that has been careful about sources and comparatively quiet about tools.\footnote{Christian Rollinger, ``Prolegomena: Problems and Perspectives of Historical Network Research and Ancient History,'' Journal of Historical Network Research 4 (2020): 1--35.}

\subsection{Three contributions}

\textbf{A body of structural observations. }We document an hourglass concentration of genealogical traffic at Leibniz; a simultaneous reorganization of seven independently extracted predicate dimensions across a window centered on the same figure; and an upstream convergence of 84.4 percent of the traced lineages on five twelfth- and thirteenth-century scholars, terminating at an eleventh-century boundary.

\textbf{A method for reading absence. }The eleventh-century cessation is, on its face, a data limitation. We argue it is also evidence. If the reason a relation went unrecorded is itself a consequence of how education was institutionally organized, then the pattern of absence and the signal of institutional change are two descriptions of one fact. We set out the conditions under which this reading is licensed, and the further work needed to discharge them.

\textbf{Tool criticism as part of the argument. }The traversal engine we built is algebraically reversible: the operations that combine attributes into a composite representation can be exactly inverted, so the reason a given candidate outranked another can be reconstructed afterward rather than assumed.\footnote{H. Chuma, K. Otsuka, and Y. Sato, ``Beyond LLMs, Sparse Distributed Memory, and Neuromorphics: A Hyper-Dimensional SRAM-CAM \emph{VaCoAl} for Ultra-High Speed, Ultra-Low Power, and Low Cost,'' arXiv:2604.11665 (2026). The companion paper reports the traversal engine and uses the same corpus as a stress test for multi-hop retrieval.} We use this property three ways: to characterize the engine's own measurement bias in closed form, to demonstrate that the macro-structures survive switching that bias off, and to report the family of lineages the traversal returns at different resolutions instead of a single list. Section 4.3 explains why reversibility is a source-critical property and not merely an engineering one.

\subsection{What is and is not claimed}

Our claims are structural descriptions of a particular graph under a particular relation. We do not estimate effects, and we do not claim that the transmission we can see is the transmission that occurred. The mentor--student predicate captures one channel among several. The Republic of Letters is so named because ideas moved by correspondence, and the twelfth-century translation movement moved them by manuscript. Where we can recover a second channel indirectly (Section 5.4) we do so and mark it clearly as indirect.

We also decline one inference the data appear to invite. In an earlier version of this work we observed that the probability of 84.4 percent convergence arising by chance is very low. We withdraw that formulation. No null model is specified for a graph whose degree distribution, temporal ordering, and coverage are all endogenous to the recording practices under study, and a significance claim without such a model is not interpretable.

\subsection{Relation to existing work}

Three literatures bear on this study.

The first is the network analysis of premodern learned institutions. De la Croix and Morault analyze the European university network across the Protestant Reformation;\footnote{David de la Croix and Pauline Morault, ``Winners and Losers from the Protestant Reformation: An Analysis of the Network of European Universities,'' Journal of Economic History 86, no. 2 (2026): 460--500.} de la Croix, Scebba, and Zanardello trace the spread of ideas through shared institutional affiliation in universities and academies, and Zanardello studies early modern academies specifically.\footnote{David de la Croix, Rossana Scebba, and Chiara Zanardello, ``Flora, Cosmos, Salvatio: Pre-modern Academic Institutions and the Spread of Ideas,'' CEPR Discussion Paper DP20569 (2025); Chiara Zanardello, ``Early Modern Academies, Universities, and Economic Growth,'' LIDAM Discussion Paper 2024/12 (2024).} These studies build from faculty lists and institutional registers, and are therefore far less exposed to the notability threshold that conditions our source, a difference we discuss in Section 3.3. Koschnick studies teacher-directed scientific change in the English Scientific Revolution, and is the closest existing work to our own question, differing chiefly in that it identifies effects within a bounded setting where we describe structure across an unbounded one.\footnote{Julius Koschnick, ``Teacher-Directed Scientific Change: The Case of the English Scientific Revolution,'' EHES Working Paper 274 (2025).} De la Croix and Goñi examine the biological counterpart of our channel.\footnote{David de la Croix and Marc Goñi, ``Nepotism vs. Intergenerational Transmission of Human Capital in Academia (1088--1800),'' Journal of Economic Growth 29, no. 4 (2024): 469--514.}

The second is the quantitative use of large biographical corpora, from cross-verified biographical databases to studies of creativity across European cities.\footnote{Morgane Laouenan et al., ``A Cross-Verified Database of Notable People, 3500BC--2018AD,'' Scientific Data 9, no. 1 (2022): 290; Michel Serafinelli and Guido Tabellini, ``Creativity over Time and Space: A Historical Analysis of European Cities,'' Journal of Economic Growth 27, no. 1 (2022): 1--43; Matias Cabello, ``The Counter-Reformation, Science, and Long-Term Growth: A Black Legend?,'' SSRN Working Paper 4389708 (2023).} On the textual side, a growing body of work recovers the movement of ideas from print rather than from persons, an approach complementary to ours and one that reaches channels a mentorship predicate cannot.\footnote{Peter Grajzl and Peter Murrell, ``A Macroscope of English Print Culture, 1530--1700, Applied to the Coevolution of Ideas on Religion, Science, and Institutions,'' Social Science History 48 (2024): 489--519; Ali Almelhem et al., ``Enlightenment Ideals and Belief in Progress in the Run-up to the Industrial Revolution: A Textual Analysis,'' Quarterly Journal of Economics 141, no. 1 (2026): 263--314.}

The third is prior work on the Mathematics Genealogy Project itself. Large-scale analyses of that dataset exist,\footnote{Sean A. Myers, Peter J. Mucha, and Mason A. Porter, ``Mathematical Genealogy and Department Prestige,'' Chaos 21, no. 4 (2011): 041104.} and our claim to novelty must be stated against them. It is not that the data have gone unanalyzed. It is that prior analyses have taken the graph as a network of departments and disciplines in the modern period, whereas we traverse it exhaustively to its documentary limit, integrate it with eleven further predicate dimensions from Wikidata, and treat the point at which it terminates as a finding.

\section{Historical Background: What a Mentor--Student Tie Was}

A single Wikidata predicate encodes, as one relation, arrangements that differed profoundly across the periods and cultures this study spans. Before any traversal is interpretable, that heterogeneity has to be stated. This section is deliberately placed before the data description, because it determines what the data can mean.

\subsection{The modern doctoral relation}

The relation most readers will supply by default, a personally supervised research dissertation culminating in a defended original contribution, is a nineteenth-century German invention, tied to the Prussian university reforms of the 1800s and 1820s and to the research-seminar model that followed. Weber documents the professionalization of the academic career that resulted.\footnote{Max Weber, ``Science as a Vocation'' (1919), in The Vocation Lectures, ed. David Owen and Tracy B. Strong (Indianapolis: Hackett, 2004).} The Mathematics Genealogy Project's core data are of exactly this kind, and they dominate the graph from roughly 1800 onward.

\subsection{The medieval and early modern doctorate}

Earlier European doctorates meant something different: a \textit{licentia docendi}, a license to teach, earned through public disputation rather than through supervised original research. The relation recorded between a candidate and a named master was one of formation and examination, not of dissertation supervision. Its social form varied by region. In the northern collegiate universities, master and students commonly lived together; in the southern student-corporation universities on the Bolognese model, the relation was contractual and the master was in a real sense the students' employee. A tie asserted between a fifteenth-century master and his student is a claim about a different social object than a tie asserted between a 1970s adviser and advisee.

\subsection{Islamic and Byzantine arrangements}

The upstream portion of our graph passes through institutions organized on other principles again. Transmission in the Islamic scholarly tradition was certified through the \textit{ij\={a}za}, a license granted by a named teacher to a named student to transmit a specific text or body of teaching, a relation both more personal and more text-specific than a European doctorate. The \textit{madrasas} of Bukhara and elsewhere predate the Western European university as systematic institutions of higher learning. Byzantine arrangements differed again. What Wikidata records for these figures is generally a ``student of'' assertion derived from the historiography, not from an institutional register.

\subsection{What follows for the analysis}

Three consequences run through the rest of the paper.

First, path length is not a homogeneous unit. A chain of twenty ties does not represent twenty instances of the same social relation, and we therefore avoid any interpretation that treats generational distance as a comparable quantity across periods.

Second, the density of recording varies with institutional form in a way that is not random. Institutions that certified individuals by name against a named master generate records; institutions that formed scholars communally do not. This is the substantive content of the Monastery Wall (Section 5.3) and the reason we treat the boundary as evidence rather than only as loss.

Third, the appropriate claim is about the graph as the sources constitute it. Where we write that a lineage passes through a figure, the claim is that the documentary record, as aggregated by these particular projects, asserts a chain of ties through that figure. Whether intellectual content traveled the same route is a separate question, on which a mentorship graph is largely silent.

\section{Sources}

\subsection{The three projects}

Our data come from Wikidata, which aggregates two specialist genealogical projects.

The Mathematics Genealogy Project (MGP), maintained at North Dakota State University, records doctoral adviser--advisee relations for mathematicians and, increasingly, adjacent fields. Its coverage of the modern period is unusually deep for a discipline-specific prosopographical resource. It is also, by its own account, substantially crowdsourced.\footnote{The Mathematics Genealogy Project states on its own website that it depends on information from its visitors for most of its data.}

The MacTutor History of Mathematics Archive, maintained at the University of St Andrews, is a curated biographical archive rather than a genealogical register. Its coverage extends much further back than that of MGP and includes many premodern and non-European figures, but it records teacher--student relations discursively, as biographical fact, rather than as a structured field.

Wikidata links to both through dedicated identifier properties (P549 for MGP, P1563 for MacTutor) and stores the relations themselves in three properties of its own: doctoral advisor (P184), student (P802), and student of (P1066). These three are not synonyms. P184 asserts the modern doctoral relation specifically. P1066 asserts the general relation ``was taught by.'' P802 is the inverse direction of P1066. We extract all three and treat them as a single relation; Section 6.4 assesses what that decision costs.

\subsection{Extraction and construction}

From the Wikidata SPARQL endpoint (snapshot accessed 2026) we extracted all entities linked by P184, P802, and P1066 within the component reachable from mathematics-oriented persons. The raw extraction contained approximately 470,000 assertions.

A pre-purification pass detected and removed 369 bidirectional pairs, that is, cases where A is asserted as the teacher of B and B as the teacher of A. These are not automatically errors. Reciprocal instruction between peers, or successive roles over a long association, are historically possible, and we note this in Section 7 as unfinished business. They are, however, incompatible with backward traversal, since they generate cycles of unbounded length. We removed them and record their count so that the decision is auditable. The result is a directed acyclic graph of 372,853 persons.

\subsection{Coverage bias and the notability threshold}

Wikidata is not a register. Inclusion requires notability by Wikimedia criteria, and the resulting coverage is skewed in several directions at once. It is skewed linguistically and geographically, toward English-, German-, and French-language scholarship. It is skewed temporally, modern records being far denser than medieval ones. It is skewed institutionally, university-affiliated practitioners being better covered than court astronomers, instrument makers, or scholars working outside recognizable institutions. And it is skewed idiomatically: the doctoral-adviser idiom travels backward more readily into European university contexts than into any other.

This last point deserves emphasis, because it is where our source differs most sharply from the closest comparable work. Studies built from faculty lists and institutional registers do not lose a scholar because that scholar was insufficiently famous; the register recorded everyone the institution employed.\footnote{De la Croix and Morault, ``Winners and Losers''; de la Croix, Scebba, and Zanardello, ``Flora, Cosmos, Salvatio.''} Wikidata does lose such scholars. Any statement we make about the structural position of a node is therefore conditional on that node having cleared a notability threshold that has nothing to do with the eleventh century and everything to do with the twenty-first.

\subsection{Crowdsourcing and representativity}

A further risk is distinct from coverage and less often addressed. MGP data are largely contributed by the community they describe, which introduces the systematic distortions that demographers have documented in online genealogies generally. Such genealogies systematically oversample elites, and a parallel literature in historical demography treats representativity as a quantity to be estimated rather than assumed.\footnote{Robert Stelter and Diego Alburez-Gutiérrez, ``Representativeness Is Crucial for Inferring Demographic Processes from Online Genealogies: Evidence from Lifespan Dynamics,'' Proceedings of the National Academy of Sciences 119, no. 10 (2022): e2120455119.}

We do not have a correction for this, and we do not claim one. What we can say is directional. Elite oversampling in a mentorship graph should, if anything, inflate the visibility of well-documented European lineages relative to others. Our most striking upstream finding, that traffic concentrates on a small set of non-European names, runs against that direction. A fuller and less elite-skewed archive could dilute the share of any particular chokepoint, but the mechanism by which it would do so is by adding alternative routes, not by removing the ones we observe.

We flag as a priority for future work the construction of a validation sample: a set of ties independently verified against institutional records, against which link precision could be estimated rather than argued.

\subsection{Variable definitions}

Every quantity used below is defined in Tab. 1.

\medskip
\noindent \textbf{Tab. 1 Variable definitions and their Wikidata sources.}

\begin{center}\small
\begin{tabular}{@{}p{0.225\linewidth}p{0.470\linewidth}p{0.235\linewidth}@{}}
\toprule
\textbf{Variable} & \textbf{Definition} & \textbf{Source property} \\
\midrule
Mentor--student tie & A directed assertion that X was taught by Y & P184, P802 (inverted), P1066 \\
Path & A sequence of ties from a tracer node backward to an ancestor & derived \\
Generation & Position of a node along a path, counted from the tracer node & derived \\
Path count & Number of distinct tracer lineages passing through a node & derived \\
Learned-society membership & Count of distinct organizations of which the person is asserted a member & P463 (member of) \\
Field & Asserted academic field & P101 (field of work) \\
Language & Asserted language used & P1412 \\
Employer & Asserted employing institution & P108 \\
Birthplace & Asserted place of birth & P19 \\
Era & Derived from birth and death dates & P569, P570 \\
Calculus affinity & Similarity of the unbound FIELD component to a composite of 26 calculus-related field tokens & derived (Section 4.3) \\
\bottomrule
\end{tabular}\end{center}

Learned-society membership therefore means the number of distinct entities linked from the person by P463. It does not distinguish an academy from a professional body, nor a full member from a corresponding member. The latter distinction matters historically, since a corresponding member of the Académie could contribute from another country whereas a university chair required residence, and we return to it as a limitation in Section 7.

Field is a coarse modern tag applied retrospectively. It loses disciplinary purity upstream, where the boundaries among mathematics, natural philosophy, and astronomy were porous or nonexistent. We use it only for relative comparison within the graph and never as a claim about what a premodern scholar considered himself to be doing.

\subsection{Summary statistics}

\medskip
\noindent \textbf{Tab. 2 Summary statistics for the extracted graph and the traversal.}

\begin{center}\small
\begin{tabular}{@{}p{0.531\linewidth}p{0.399\linewidth}@{}}
\toprule
\textbf{Quantity} & \textbf{Value} \\
\midrule
Raw mentor--student assertions extracted & approx. 470,000 \\
Bidirectional pairs removed & 369 \\
Persons in the resulting DAG & 372,853 \\
Tracer set & 64 Fields Medalists (63 with a recorded mentor) \\
Distinct paths enumerated (FS = 20,000) & approx. 25.5 million \\
Maximum generational depth reached & 57 \\
Path-to-node ratio & approx. 68:1 \\
Hub nodes enriched with further predicates & 1,043 \\
Edges among enriched hub nodes & 1,408 \\
Predicate dimensions attached per hub node & up to 12 \\
Nodes at the maximum upstream path count (54) & 213 \\
\bottomrule
\end{tabular}\end{center}

\subsection{The tracer set}

We use the complete set of historical Fields Medalists (64 laureates) solely as a tracer set: a uniform, ex ante criterion for which present-day lineages to follow backward. The Medal is not a claim about the primacy of mathematics. A fixed, publicly known starting set makes the exercise replicable, since any reader can recover which nodes seed the traversal, and the findings concern the shape of the induced subgraph rather than the merits of the seeds.

The choice does have consequences, and Section 6.2 assesses them. The most obvious is that it privileges lineages with a mathematical thread, which upstream becomes a thread through the quadrivium and its Islamic and Byzantine antecedents. A different tracer set would change absolute counts and might change the identities of the chokepoints. What it would not change is the question the exercise makes well posed: on a graph too large for unaided inspection, where does the topology concentrate?

\section{Method and Tool Criticism}

Digital hermeneutics distinguishes four kinds of criticism: of the source, of the tool, of the algorithm, and of the interface. Section 3 was source criticism. This section is tool and algorithm criticism, and we treat it as load-bearing rather than as an appendix, because the traversal makes choices that materially determine which lineages the reader is shown.

\subsection{Why the traversal is hard}

Academic genealogy on this graph is not a tree. A scholar may have several recorded teachers; any teacher typically has many students; common ancestors are reached through overlapping descendant lineages. The structure is a directed acyclic graph, and the number of distinct paths from a source to its ancestors grows multiplicatively with depth even when the number of distinct nodes is modest. From 64 sources on 372,853 nodes, the path count reaches 25.5 million at depth 57, roughly 68 paths per node.

Exhaustive enumeration at this scale is not feasible, and three standard alternatives fail for structural rather than incidental reasons. A hash-table traversal exhausted 128 GB of main memory at a frontier size of 25,000 and became unexecutable; it also, when it does run, orders candidates within the frontier by the lexicographic order of their identifiers, an ordering with no historical content whatever. A GPU-based hyperdimensional library requested over 100 GB of video memory for the semantic stage and terminated with an out-of-memory error on a 12 GB card. Standard centralities such as PageRank and betweenness complete in reasonable time but are field-blind: they cannot distinguish a mathematician from an anatomist occupying the same topological position. Section 6.3 shows that this is not a hypothetical concern on this corpus.

\subsection{Frontier size and what completeness means}

The traversal is therefore executed under a frontier size (FS) cap, an upper bound on the number of concurrent candidate paths retained at each generation. We set FS = 20,000 for the main analysis.

This has an implication worth stating plainly, because it applies to any method operating on a combinatorially explosive DAG and not only to ours: there is no absolute completeness to be had. The exactness of a hash table is exactness within the frontier, and nothing beyond. FS relativity is a property of the problem, not a limitation of one instrument. What differs between instruments is therefore not whether they prune, but how they choose what to keep. This is the question tool criticism has to answer, and it is where the reversibility of our engine does real work.

\subsection{Reversibility as a source-critical property}

The engine represents each person as a very high-dimensional binary vector, and represents attributes by combining such vectors with two operations: binding, which attaches a value to a role, for example FIELD to astronomy, and bundling, which superposes several bound pairs into one composite. Similarity between vectors is measured by Hamming distance. For uncorrelated vectors in a binary space the expected similarity is 0.5, and this value serves as the baseline throughout.

The property that matters here is that binding is exactly invertible. It is implemented as XOR-and-shift over the binary field GF(2), and applying the operation twice returns the original. One can therefore unbind a composite, recovering the FIELD component of the representation of a person separately from the LANGUAGE or EMPLOYER components, without decompressing or re-encoding anything, and without loss.

Three reasons this is a source-critical rather than an engineering point.

It makes the ranking auditable. The engine attaches to every retained path a cumulative score, below CR2, that records how cleanly each step along that path was resolved. Because the algebra is reversible and the diffusion deterministic, one can go back to a branch point, take a different edge, and compare, rather than merely observe that a different answer emerged. A historian asking why this lineage surfaced and that one did not gets an answer that can be checked, not a black-box assertion.

It permits multi-predicate measurement. A hash lookup answers presence or absence on a single key. It has no operation corresponding to ``extract the FIELD component of this composite representation and measure how far it lies from calculus.'' Reversibility is what turns a lookup structure into a measuring instrument, and Sections 5.2 and 5.4 depend entirely on it.

It bounds the distortion of the tool itself in closed form. This is the subject of the next subsection. We give the full algebraic specification elsewhere and summarize it in Appendix B.\footnote{Full specification in Chuma, Otsuka, and Sato, arXiv:2604.11665.} A reader interested only in the historical findings needs one fact from this subsection: the measurement error of the engine is characterized exactly, and is three orders of magnitude smaller than the effects reported in Section 5.

\subsection{The bias of the tool, characterized}

The engine partitions each representation across N independent blocks, each of which votes for a candidate. When two distinct items map to the same block address, that block is marked Don't Care and its vote is withheld rather than being allowed to produce a false majority. In single-step retrieval this is straightforward noise tolerance. Under multi-step traversal it acquires a second-order effect we did not design and did not anticipate. Configurations are written below in the form N\_m, so that the main analysis configuration of N = 128 blocks at memory-depth exponent m = 27 appears as 128\_27 in Figs. 1 and 2.

Write CR1 for the per-step agreement rate among blocks, and CR2, which we also call the cumulative path integral, for its running product along a path, so that CR2(n) = CR2(n$-$1) $\times$ CR1(n$-$1). In the main configuration CR1 stabilizes just below unity, measured at approximately 0.997, so single steps are near-perfect. But the product compounds. Two quantities must be kept apart here, because both appear in the figures. The closed-form prediction is CR1 raised to the power n, which at n = 56 gives 0.846. The measured trajectory across all enumerated paths reaches 0.905 at the same depth. The gap between the two is consistent with survivorship along deep paths, since paths whose CR2 fell furthest were pruned before reaching that depth.

Fig. 1 plots the measured trajectory, in the 128\_27 configuration and against the 128\_10 configuration with rescue engaged, and annotates its endpoint of 0.905. Fig. 2, which sweeps the resolution parameter, plots the closed-form curves instead: its N = 128 endpoint of 0.85 is therefore the prediction and not the measurement, and the two figures are consistent once that is understood. We give the closed-form values in Tab. 3 for the same reason, so that the table and the lower panel of Fig. 2 can be read against each other directly.

The consequence for ranking is that short, direct routes accumulate less penalty than long, circuitous ones, and the traversal prefers them accordingly. This is a parsimony criterion the architecture imposes on itself. We report it because it is exactly the kind of thing tool criticism exists to catch. An instrument that produces path-dependent output, applied to a question about historical path dependence, is a circularity worth taking seriously. Section 6.3 addresses it directly. Fig. 1 shows the two trajectories side by side.

\begin{figure}[htbp]\centering
\includegraphics[width=\linewidth]{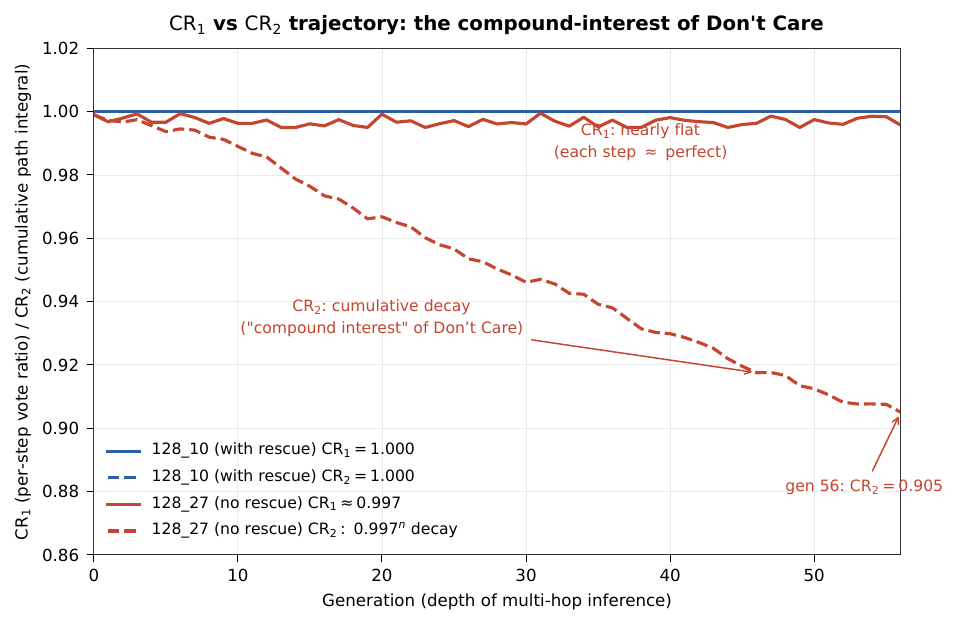}
\caption*{\footnotesize \textbf{Fig. 1 Generational trajectories of CR1 and CR2.} The 128\_27 configuration (no rescue; main analysis) against 128\_10 (with rescue). In 128\_27 the per-generation CR1 stays within 0.995--0.999, so single-step retrieval is near-perfect, while CR2 decays monotonically, reaching 0.905 at the 56th generation. With rescue, both are pinned at 1.000, suppressing the selection mechanism entirely.}
\vspace{-0.4em}{\scriptsize Figure by the authors, plotted from the per-generation confidence logs of the traversal engine in the 128\_27 and 128\_10 configurations. A version of this figure appeared in Chuma, Otsuka, and Sato, arXiv:2605.31470; copyright is retained by the authors. The plotted values are deposited with the supplementary data.}
\end{figure}

\subsection{Resolution as part of the result}

The engine has a resolution parameter: the number of blocks N, traded against the addressable depth per block under a fixed total capacity. Sweeping it does not produce better and worse approximations to one answer. It produces a family of internally coherent macro-structures (Tab. 3 and Fig. 2).

\medskip
\noindent \textbf{Tab. 3 Regimes and lineages surfaced at each resolution, under the fixed total-capacity constraint.} Collision rate is the count-based per-block rate; depth is the maximum generation reached. The CR2 column gives the closed-form endpoint CR1 raised to the power 57, which is the value plotted and annotated in the lower panel of Fig. 2. Regime names are those used in the legend of that figure.

\begin{center}\small
\begin{tabular}{@{}p{0.216\linewidth}p{0.119\linewidth}p{0.087\linewidth}p{0.119\linewidth}p{0.389\linewidth}@{}}
\toprule
\textbf{Configuration} & \textbf{Collision rate} & \textbf{Depth} & \textbf{CR2 at gen. 57} & \textbf{Regime and lineage surfaced} \\
\midrule
N = 64, m = 28 & 0.074\% & 55 & 0.91 & Chimera: fragmented superposition, central hub lost \\
N = 128, m = 27 & 0.144\% & 55 & 0.85 & Transitional: Leibniz--Euler line, with residual Gauss-lineage crossover \\
N = 256, m = 26 & 0.282\% & 55 & 0.63 & Gauss alternative: the Gauss line (Pfaff, Hausen, Kästner) \\
N = 512, m = 25 & 0.559\% & 57 & 0.42 & Orthogonal purification: Leibniz--Euler line, purified through Lagrange and Poisson \\
N = 1024, m = 24 & 1.108\% & 56 & 0.21 & Capacity boundary: Leibniz-dominant, with partial Gauss re-entry \\
\bottomrule
\end{tabular}\end{center}

\begin{figure}[htbp]\centering
\includegraphics[width=\linewidth]{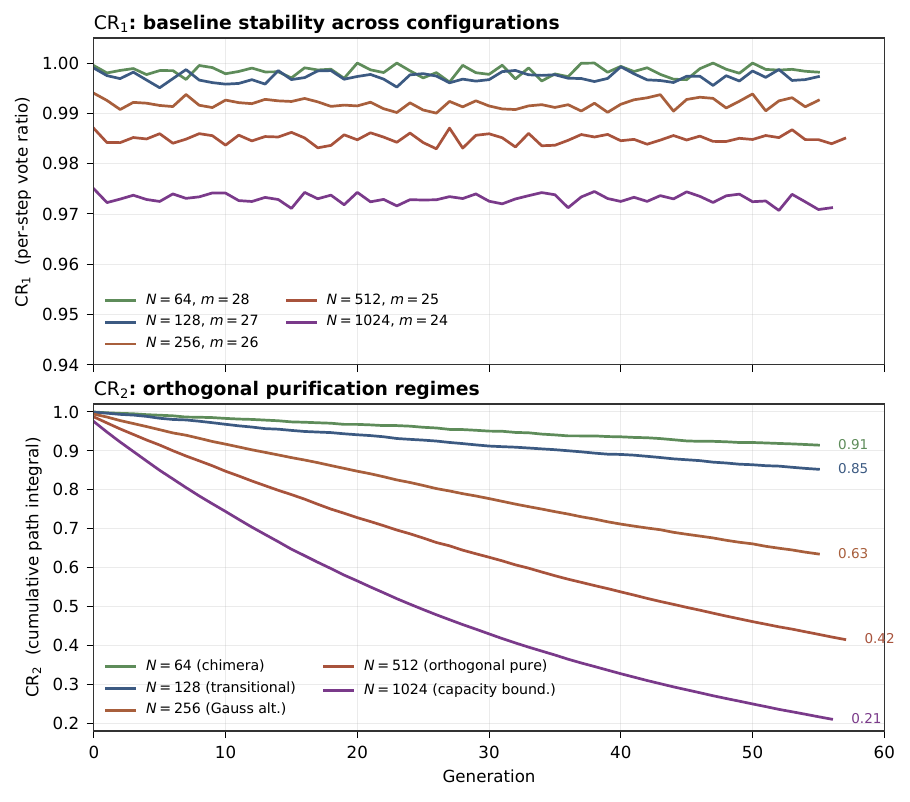}
\caption*{\footnotesize \textbf{Fig. 2 CR1 and CR2 across the five resolutions.} Upper band: CR1 stays near 1.0 regardless of dimensionality, confirming that the macroscopic signal is never lost. Lower trajectories: CR2 decays at rates depending on N. Low-N configurations remain persistently high, the chimera regime; N = 512 shows a steep yet smooth decay, orthogonal purification, and reaches the maximum depth of 57 generations; N = 1024 shows the steepest decay, the capacity boundary under the fixed total-capacity constraint.}
\vspace{-0.4em}{\scriptsize Figure by the authors, plotted from the confidence logs of the five-configuration resolution sweep under the fixed total-capacity constraint. A version of this figure appeared in Chuma, Otsuka, and Sato, arXiv:2605.31470; copyright is retained by the authors. The plotted values are given in Tab. 3 and deposited with the supplementary data.}
\end{figure}

Two things are worth noting. Higher collision rates do not degrade the result: the fourfold rise from N = 128 to N = 512 improves the coherence of the extracted lineage. And the Gauss line surfaced at N = 256 is not an error. It is a mutually exclusive but historically real macro-structure, one that a traversal returning a single ranked list would simply have concealed.

We therefore report the family and the parameter that indexes it. Reporting that at N = 256 the Gauss lineage dominates while at N = 512 the Leibniz--Euler lineage dominates, and that both are historically defensible, is a more informative act than reporting either alone. The resolution parameter is not noise to be tuned away before publication. It is part of what was found.

This holds, we should say, because the mentor--student graph is a DAG with strong temporal ordering, so that the cumulative penalty correlates cleanly with generational distance. How resolution families behave on ontologies with different topologies, bidirectionally cyclic ones in particular, is open.

\subsection{Provenance of reported figures}

Because the preceding two subsections establish that different configurations produce different outputs, every number in this paper must carry its provenance. Tab. 4 supplies it, and we ask readers to treat any figure not traceable there as unverified.

\medskip
\noindent \textbf{Tab. 4 Provenance of every figure reported in this paper.} RR denotes the rescue rate: RR = 1 repairs all collisions, RR = 0 tolerates them.

\begin{center}\small
\begin{tabular}{@{}p{0.310\linewidth}p{0.441\linewidth}p{0.179\linewidth}@{}}
\toprule
\textbf{Figure reported} & \textbf{Configuration} & \textbf{Mode} \\
\midrule
25.5 million paths; depth 57 & Hash-table baseline at FS = 20,000, reproduced bitwise by the engine & Rescue (RR = 1) \\
25.43 million paths; 1,043 nodes; 1,408 edges & N = 128, m = 27 & Don't Care (RR = 0) \\
CR1 $\approx$ 0.997; CR2 $\approx$ 0.905 at generation 56 & N = 128, m = 27 & Don't Care \\
Hourglass ratios (Tab. 5) & Invariant across configurations & Both \\
Seven-indicator reorganization (Tab. 7) & N = 128, m = 27; three continuity indicators strengthen at N = 512 & Don't Care \\
84.4 percent upstream convergence & FS = 20,000; stable for FS $\geq$ 10,000 & Both \\
Within-frontier rankings (Tab. 12) & As labeled per column & As labeled \\
\bottomrule
\end{tabular}\end{center}

The distinction in the first two rows is the one most easily lost. The engine in Rescue mode repairs every collision and reproduces the hash-table baseline exactly, yielding 25.5 million paths. In Don't Care mode it tolerates collisions, applies the parsimony penalty of Section 4.4, and consequently culls some deep paths, about 70,000 of them. Both figures are correct; they answer different questions. Where we quote the scale of the graph we use the former; where we discuss what the traversal selected we use the latter.

\section{Findings}

\subsection{An hourglass at Leibniz}

Genealogical traffic centered on Leibniz forms an hourglass: a narrow band of midstream positions concentrates traffic, while wide sets of more distant nodes feed in from the past and diffuse out toward the present. Upstream, nodes on Leibniz-traversing paths carry an average of 5.3 paths; downstream the figure expands to 53.4, a thickness ratio of 10.1 (Tab. 5). Fig. 3 shows the profile against that of Newton, drawn to the same scale.

\medskip
\noindent \textbf{Tab. 5 Path counts and thickness ratios at hub nodes.} Ratio is the student-direction mean divided by the mentor-direction mean.

\begin{center}\small
\begin{tabular}{@{}p{0.161\linewidth}p{0.322\linewidth}p{0.322\linewidth}p{0.124\linewidth}@{}}
\toprule
\textbf{Hub} & \textbf{Mean paths, mentor direction} & \textbf{Mean paths, student direction} & \textbf{Ratio} \\
\midrule
Leibniz & 5.3 & 53.4 & 10.1 \\
Gauss & 2.7 & 41.2 & 15.1 \\
Hilbert & 1.6 & 29.2 & 18.5 \\
Euler & 3.6 & 52.4 & 14.6 \\
Newton & approx. 1 & approx. 2 & -- \\
\bottomrule
\end{tabular}\end{center}

\begin{figure}[htbp]\centering
\includegraphics[width=\linewidth]{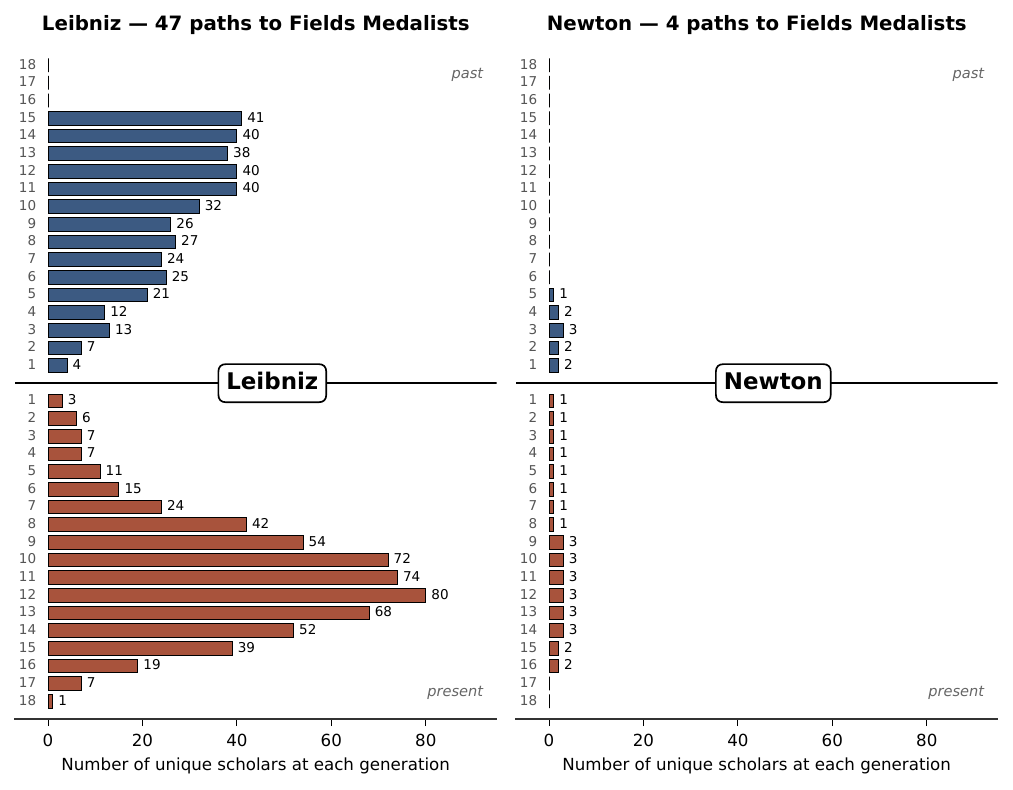}
\caption*{\footnotesize \textbf{Fig. 3 Hourglass topology of genealogical traffic: Leibniz and Newton, drawn to the same scale.} Each bar is the number of unique scholars at a given generational distance through which paths to Fields Medalists pass. Upper half: mentor direction, toward the past. Lower half: student direction, toward the present. Leibniz shows the constriction-and-expansion profile; Newton is a thin thread with no hub structure in either direction.}
\vspace{-0.4em}{\scriptsize Figure by the authors, plotted from the enumerated path set on the Wikidata mentor--student graph at FS = 20,000. A version of this figure appeared in Chuma, Otsuka, and Sato, arXiv:2605.31470; copyright is retained by the authors. The underlying Wikidata content is available under CC0, and the per-generation counts are deposited with the supplementary data.}
\end{figure}

Fig. 3 gives the underlying counts. On the Leibniz side the mentor direction rises to 41 unique scholars at the fifteenth generation upstream, while the student direction peaks at 80 at the twelfth generation downstream, then tapers to a single scholar at the eighteenth. The profile of Newton is the contrast. Only four of the 64 lineages pass through him; his mentor direction reaches at most three unique scholars and stops by the fifth generation, and his student direction never exceeds three at any depth. The node shows no constriction and expansion in either direction: a thin thread rather than an hourglass.

The comparison requires care, and we want to be explicit about what it is not. It is not a valuation of one national scientific culture over another, and it is not a claim about intellectual merit. It is a property of a directed graph built from a single predicate, and the two figures are compared because they carry comparable weight in the history of the calculus while differing sharply in the depth and breadth of documented lineage running through them.

The documented difference has a documented cause. Leibniz founded the Berlin Academy of Sciences in 1700, maintained one of the largest scholarly correspondence networks in Europe, and stands at the head of a lineage running through Johann Bernoulli, Euler, Lagrange, Laplace, and Poisson. Newton trained few successors. This is not a mystery requiring a network to reveal it; what the network adds is a measure of the magnitude.

A note on the priority dispute. The three-century quarrel over the calculus turned on a question, who was first, to which a genealogy can add nothing.\footnote{A. Rupert Hall, Philosophers at War: The Quarrel between Newton and Leibniz (Cambridge: Cambridge University Press, 1980). On the campaign subsequently conducted against Leibniz, see Maria Rosa Antognazza, Leibniz: A Very Short Introduction (Oxford: Oxford University Press, 2016).} It can, however, distinguish two things the dispute conflated: possession of a body of technique and transmission of it. We construct a composite score combining the calculus affinity of a node (Section 4.3) with its normalized path count, such that only nodes high on both rank highly (Tab. 6).

\medskip
\noindent \textbf{Tab. 6 Composite score combining semantic affinity with structural position.} Calculus affinity is the Hamming similarity between the node's unbound FIELD component and a composite of 26 calculus-related field tokens, rescaled from the 0.5--0.6 range onto 0--1; the baseline for uncorrelated vectors is 0.5. Paths is the number of distinct tracer lineages through the node. Composite is the product of rescaled affinity and path count normalized by the network maximum. Ranks are among all 1,043 enriched hub nodes.

\begin{center}\small
\begin{tabular}{@{}p{0.103\linewidth}p{0.323\linewidth}p{0.181\linewidth}p{0.207\linewidth}p{0.116\linewidth}@{}}
\toprule
\textbf{Rank} & \textbf{Scholar (dates)} & \textbf{Composite} & \textbf{Calculus affinity} & \textbf{Paths} \\
\midrule
1 & J. F. Pfaff (1765--1825) & 0.841 & 0.638 & 33 \\
2 & Leibniz (1646--1716) & 0.597 & 0.569 & 47 \\
3 & J. Bernoulli (1655--1705) & 0.423 & 0.551 & 45 \\
12 & Euler (1707--1783) & 0.244 & 0.529 & 45 \\
17 & Gauss (1777--1855) & 0.167 & 0.531 & 29 \\
91 & Newton (1643--1727) & 0.025 & 0.534 & 4 \\
\bottomrule
\end{tabular}\end{center}

The calculus affinity of Newton, 0.534, is almost indistinguishable from that of Leibniz, 0.569. His rank of 91 is driven entirely by a path count of four. Read as a statement about the sources, this says that the record documents Newton as possessing the calculus and Leibniz as both possessing and transmitting it, a distinction long latent in accounts of how useful knowledge became cumulative.\footnote{Joel Mokyr, A Culture of Growth: The Origins of the Modern Economy (Princeton: Princeton University Press, 2016); Mara P. Squicciarini and Nico Voigtländer, ``Human Capital and Industrialization: Evidence from the Age of Enlightenment,'' Quarterly Journal of Economics 130, no. 4 (2015): 1825--1883.} Whether it reflects the historical reality of transmission, or only the reality of what got written down, is precisely the question Section 6 takes up.

\subsection{A multi-predicate reorganization around 1700}

If the seventeenth-century concentration reflects a durable change in transmission infrastructure rather than a documentary accident, the change should appear across independent attributes and not only in path counts. We test this using the reversibility described in Section 4.3: for each of the 1,043 enriched hub nodes we bind up to twelve predicate dimensions into one composite representation, then unbind each dimension in turn and measure it separately.

Seven indicators result (Tab. 7). Their significance lies in their independence. Each is extracted by a separate unbinding operation from a different predicate dimension, so a single indicator could be dismissed as coincidence or as a documentary artifact, whereas simultaneous movement across seven cannot be so easily.

\medskip
\noindent \textbf{Tab. 7 Indicators of reorganization across the Leibniz window.} The three continuity indicators measure compositional overlap between the pre- and post-windows on a scale where 0.5 denotes no change; values above 0.5 indicate continuous transformation rather than replacement. Diffusion entropy is Shannon entropy across the field, language, and institution dimensions jointly. Top-tier production rate is the proportion of Leibniz's direct students whose descendants reach the highest decile of path count. Windows are defined in Tab. 8. Values are for N = 128; the three continuity indicators strengthen at N = 512.

\begin{center}\small
\begin{tabular}{@{}p{0.644\linewidth}p{0.286\linewidth}@{}}
\toprule
\textbf{Indicator} & \textbf{Value} \\
\midrule
Learned-society membership, mean per person (pre to post) & 0.08 to 3.81 (46-fold) \\
Field continuity & 0.614 \\
Language continuity & 0.603 \\
Employer continuity & 0.603 \\
Institutional hub diversification (post/pre) & 2.5-fold \\
Direct-student top-tier production rate & 1.000 \\
Influence diffusion entropy & 3.401 \\
\bottomrule
\end{tabular}\end{center}

The strongest single indicator is learned-society membership. Tab. 8 gives the underlying cross-section.

\medskip
\noindent \textbf{Tab. 8 Learned-society membership by window.} Windows are defined by Leibniz's birth and death years and applied to birth-year overlap. Cohorts comprise the enriched hub nodes only. Membership counts distinct P463 links.

\begin{center}\small
\begin{tabular}{@{}p{0.203\linewidth}p{0.119\linewidth}p{0.250\linewidth}p{0.191\linewidth}p{0.167\linewidth}@{}}
\toprule
\textbf{Window} & \textbf{Scholars} & \textbf{With recorded membership} & \textbf{Total memberships} & \textbf{Mean per person} \\
\midrule
Pre (to 1646) & 434 & 28 (6.5\%) & 36 & 0.08 \\
Interim (1646--1716) & 87 & 44 (53.0\%) & 107 & 1.29 \\
Post (from 1716) & 504 & 414 (82.1\%) & 1,920 & 3.81 \\
\bottomrule
\end{tabular}\end{center}

The three continuity indicators all exceed the no-change baseline of 0.5, indicating transformation that builds on rather than replaces what came before. Alongside them, entropy measurement of academic languages records a standardization of the medium: from the diverse linguistic distribution of the medieval cohort (entropy 3.263), through convergence on Latin and German, to eventual consolidation around English.

Two cautions about what the reader should not conclude.

First, this is a documentary series, and the period in question is one in which documentation itself improved. A 46-fold rise in recorded memberships is compatible with a smaller rise in actual memberships plus better recording. Our defense is the joint movement: better recording of societies would not by itself produce simultaneous movement in field, language, employer, institutional diversity, and diffusion entropy, each extracted independently.

Second, and this is a fair criticism of the earlier version of this work, we compare windows defined by Leibniz because Leibniz is where the path traffic concentrates, not because we have established that Leibniz caused anything. Scholarly activity in general was changing dramatically in the seventeenth century, and pre- and post-windows around other contemporaneous nodes may well show similar patterns. Establishing whether the Leibniz window is special, or merely one sample of a general seventeenth-century shift, requires a systematic comparison across candidate pivot nodes. We have not done it, and we flag it as the single most useful robustness exercise a reader could perform on our deposited data.

\subsection{Upstream convergence and the Monastery Wall}

Turning upstream: among the 213 individuals carrying the maximum path count of 54, more than Leibniz, five twelfth- and thirteenth-century scholars emerge as the furthest-upstream hubs of the entire traced network (Tab. 9).

\medskip
\noindent \textbf{Tab. 9 The five upstream hubs, in order of mentor-to-student succession.} All five lie on the chain through which 54 of the 64 tracer lineages pass.

\begin{center}\small
\begin{tabular}{@{}p{0.103\linewidth}p{0.362\linewidth}p{0.349\linewidth}p{0.116\linewidth}@{}}
\toprule
\textbf{Order} & \textbf{Scholar} & \textbf{Dates; birthplace} & \textbf{Paths} \\
\midrule
1 & Sharaf al-D\={\i}n al-\d{T}\={u}s\={\i} & 1135--1201; Tus (Iran) & 54 \\
2 & Kam\={a}l al-D\={\i}n ibn Y\={u}nus & 1156--1242; Mosul & 54 \\
3 & Na\d{s}\={\i}r al-D\={\i}n al-\d{T}\={u}s\={\i} & 1201--1274; Tus (Iran) & 54 \\
4 & Shams al-D\={\i}n al-Bukh\={a}r\={\i} & 1254--1300; Bukhara & 54 \\
5 & Grigorios Choniades & 1240--1320; Constantinople & 54 \\
\bottomrule
\end{tabular}\end{center}

The chain terminates with Choniades, a Byzantine scholar who studied in Persia and carried Islamic astronomical and mathematical knowledge back to the Greek-speaking world.

That 54 of 64 tracer lineages, or 84.4 percent at FS = 20,000, run through five individuals constituting less than 0.001 percent of the node set is the most arresting number in this paper, and also the one most in need of qualification. Stated carefully, the claim is that on the only relation we observe at scale, and as the sources currently constitute it, a very small slice of the node set carries an overwhelming share of the paths on which the tracer set depends. It is not a counterfactual claim about causation, and, as stated in Section 1.2, it is not a significance claim. Section 6.1 assesses whether it is an artifact. The finding is consistent with existing qualitative accounts of the transmission of Greek and Arabic learning, to which it adds a structural dimension.\footnote{George Saliba, Islamic Science and the Making of the European Renaissance (Cambridge [MA]: MIT Press, 2007); Shuntaro Itoh, The Twelfth-Century Renaissance: The World of Arabic Civilisation and Western Europe (Tokyo: Iwanami Shoten, 1993) [in Japanese].}

Why the chain stops. Despite exhaustive parameter exploration, no lineage extends beyond the 57th generation. Every path terminates at one of two figures who predate the five by roughly a century: Lanfranc of Canterbury (ca. 1005--1089), who taught at the Abbey of Bec, or Ivo of Chartres (ca. 1040--1116), who taught at the cathedral school there. Neither has a recorded teacher in Wikidata.\footnote{C. Stephen Jaeger, The Envy of Angels: Cathedral Schools and Social Ideals in Medieval Europe, 950--1200 (Philadelphia: University of Pennsylvania Press, 2000). That Lanfranc taught at the Abbey of Bec and Ivo at the cathedral school of Chartres is documented; neither has a recorded teacher of his own.} We name this systematic cessation the Monastery Wall. The chronological profile of the 213 maximum-path scholars makes the boundary visible at the population level (Tab. 10).

\medskip
\noindent \textbf{Tab. 10 The 213 maximum-path scholars by century.}

\begin{center}\small
\begin{tabular}{@{}p{0.484\linewidth}p{0.223\linewidth}p{0.223\linewidth}@{}}
\toprule
\textbf{Century} & \textbf{Count} & \textbf{Share} \\
\midrule
Eleventh & 1 & 0.5\% \\
Twelfth & 13 & 6.1\% \\
Thirteenth & 21 & 9.9\% \\
Fourteenth & 24 & 11.3\% \\
Fifteenth & 34 & 16.0\% \\
Sixteenth & 79 & 37.1\% \\
Seventeenth & 39 & 18.3\% \\
Dates unknown & 2 & 0.9\% \\
Total & 213 & 100\% \\
\bottomrule
\end{tabular}\end{center}

A single eleventh-century entry; a ramp through the scholastic and Renaissance centuries; a peak in the sixteenth; a seventeenth-century tail after which Leibniz alone carries the maximum.

Absence as evidence. Two readings of the cessation compete in principle. The first is that it is structural data loss, rooted in the informality of teaching relations in monastic education. The second is that it is the detection, as network topology, of a change in the medium through which knowledge propagated, from communal monastic formation to personal academic lineage.

These are not mutually exclusive, and their relation is the methodological point. As the standard histories of the medieval university document, education in the tenth and eleventh centuries revolved around \textit{scholae monasticae} and cathedral schools, where the monastic ideal of \textit{stabilitas loci} made formation a communal affair oriented toward scriptural understanding.\footnote{Hastings Rashdall, The Universities of Europe in the Middle Ages, 3 vols. (1895; repr. Cambridge: Cambridge University Press, 2010); Hilde de Ridder-Symoens, ed., A History of the University in Europe, vol. 1, Universities in the Middle Ages (Cambridge: Cambridge University Press, 1992).} Knowledge passed through oral tradition, joint manuscript copying, and the tacit pedagogy of liturgy, channels that by their nature generate no structured record of who taught whom.

If the reason a relation went unrecorded is itself a consequence of how education was institutionally organized, then the pattern of absence is the signal of institutional change, described from the other side. This is the methodological stance we adopt: not only what can be seen, but where vision fails, is interpretable.

We are equally clear about what would be required to discharge the reading, because it is not self-validating. One would need to cross-reference the recording pattern of Wikidata against independent sources, namely monastic and cathedral-school records and the substantial historiography of eleventh-century education, and show that the shape of the absence matches the institutional account rather than merely being compatible with it. We have not done this. Until it is done, the Monastery Wall should be read as a well-motivated hypothesis about the record rather than an established fact about the past.

\subsection{Recovering a second channel}

The mentor--student predicate captures one channel of transmission. The twelfth-century translation movement, in which Arabic renderings of Greek originals and original Arabic works were translated into Latin at Toledo, in Sicily, and in northern Italy, moved knowledge through books and translators rather than through personal instruction, and is by construction invisible to a mentorship genealogy.\footnote{Itoh, Twelfth-Century Renaissance. Translation proceeded across four centers: the Aragonese school, the Toledo school, the Sicilian school, and the north Italian school.} Figures of the Toledo school such as Gerard of Cremona (d. 1187) and Michael Scot (d. ca. 1232) appear in our extracted data, but their recorded lineages do not connect to the tracer set.

It is not, however, invisible to the semantic side of the representation. Using the unbinding operation to extract the FIELD component from the composite vector of each node, and bundling across all scholars active in each 50-year window from 1100 to 1800, we obtain a per-window field signal (Tab. 11).

\medskip
\noindent \textbf{Tab. 11 Rate of field-signal change between adjacent 50-year windows.} Signal is the mean bitwise agreement between the unbound FIELD component of the bundled window representation and the field-specific reference vector; the baseline for uncorrelated vectors is 0.5. Windows are assigned by birth-year overlap.

\begin{center}\small
\begin{tabular}{@{}p{0.211\linewidth}p{0.169\linewidth}p{0.183\linewidth}p{0.183\linewidth}p{0.183\linewidth}@{}}
\toprule
\textbf{Window} & \textbf{Algebra} & \textbf{Change} & \textbf{Astronomy} & \textbf{Change} \\
\midrule
1100--1150 & 0.4999 & base & 0.5003 & base \\
1150--1200 & 0.5001 & +0.0001 & 0.4999 & $-$0.0003 \\
1200--1250 & 0.4998 & $-$0.0003 & 0.4999 & approx. 0 \\
1250--1300 & 0.4997 & $-$0.0001 & 0.5024 & +0.0025 \\
1300--1350 & 0.5001 & +0.0004 & 0.5021 & $-$0.0003 \\
1350--1400 & 0.5000 & $-$0.0001 & 0.4994 & $-$0.0027 \\
1400--1450 & 0.4995 & $-$0.0004 & 0.4987 & $-$0.0007 \\
1450--1500 & 0.4994 & $-$0.0001 & 0.5052 & +0.0065 \\
1500--1550 & 0.5003 & +0.0009 & 0.5033 & $-$0.0019 \\
1550--1600 & 0.4996 & $-$0.0007 & 0.5039 & +0.0006 \\
1600--1650 & 0.5001 & +0.0005 & 0.5076 & +0.0037 \\
1650--1700 & 0.5019 & +0.0018 & 0.5053 & $-$0.0023 \\
1700--1750 & 0.5006 & $-$0.0013 & 0.5050 & $-$0.0002 \\
1750--1800 & 0.5010 & +0.0004 & 0.5079 & +0.0029 \\
\bottomrule
\end{tabular}\end{center}

Astronomy departs upward in two windows: 1250--1300, aligning with the second Toledo translation period and the peak of the Maragha observatory school, and 1450--1500, coinciding with the lifetime of Copernicus. Algebra remains within 0.001 of baseline throughout. In the same period, unbinding of the birthplace dimension shows the geographic center of gravity migrating from the Abbasid sphere toward the Holy Roman Empire. The asymmetry between the two fields is what makes the exercise informative: a field with no signal would wobble near 0.5 in both directions, whereas astronomy departs in the same direction in windows where translation activity and Copernican-era work plausibly raised the salience of instruments, tables, and geometrical models.\footnote{Saliba, Islamic Science, on Copernicus's borrowing of the models of al-`Ur\d{d}\={\i}, al-\d{T}\={u}s\={\i}, and Ibn al-Sh\={a}\d{t}ir.}

The limits of this exercise are severe and we state them. The magnitudes are in the third and fourth decimal place. What the field signal registers is the composition of recorded field attributions among scholars active in a window; it is a measure of the changing description the archive gives of its subjects, and only derivatively of the subjects. It does not show that the intellectual inheritance of any individual passed through a Toledo manuscript. It shows that the semantic centroid of the recorded population tilts, in windows where an independent historiography says it should. We offer it as corroboration of an existing account, not as evidence that would stand alone. Textual approaches reach this channel far more directly, and a serious treatment of the translation movement should use them.\footnote{Grajzl and Murrell, ``Macroscope''; Almelhem et al., ``Enlightenment Ideals.''}

\section{Source Criticism of the Findings}

This section takes the four most serious objections to the preceding results and works through each. We regard it, rather than Section 5, as the center of gravity of the paper.

\subsection{Is the upstream convergence an artifact of missing data?}

The objection. In an ascending genealogy, every upward path must eventually break, because at some depth the record simply stops. If records thin out as one moves backward, then all surviving paths will necessarily funnel through whichever few nodes happen to retain an upward edge. Convergence on a small set is then a construction artifact, and the identity of the five tells us about the medieval coverage of Wikidata rather than about the transmission of knowledge.

What we concede. The objection is correct that convergence of some kind is structurally guaranteed. Any ascending traversal of a thinning record produces a narrowing. The bare fact that 84 percent of lineages share upstream nodes is therefore not, on its own, evidence of anything historical.

What survives. Three things. First, the identity of the funnel is not free. That some funnel exists is guaranteed; that it runs through five specific scholars in Tus, Mosul, Bukhara, and Constantinople is not. A thinning-record model predicts convergence on whichever nodes retain edges, which, given the documented European skew of Wikidata, one would expect to be European. The observed funnel runs against the coverage bias, not with it.

Second, the wall and the funnel are one century apart, not coincident. If convergence were purely the thinning artifact, we would expect the funnel and the termination to be the same event: paths would narrow and stop together. They do not. Paths narrow onto the five in the twelfth and thirteenth centuries, then continue through the eleventh century to Lanfranc and Ivo, and stop there. The hundred-year gap between the bottleneck and the wall is not what the artifact story predicts.

Third, the convergence is stable, not marginal. Across frontier sizes the figure rises from about 72 percent at FS = 5,000 to 84 percent at FS of 10,000 or more, and plateaus. A quantity determined by an arbitrary pruning limit would not plateau.

What would settle it. A proper test would construct a null model: rewire the graph preserving the degree sequence and the temporal ordering, and ask how often a randomly rewired graph produces convergence on five nodes as far upstream as these. We have not built this model, since degree-preserving rewiring on a DAG with hard temporal constraints is nontrivial and the choice of what to preserve is itself a substantive assumption. We consider it the highest-value next step, and we flag that until it exists, the 84 percent figure should be read as a description of the traced subgraph and not as a claim about improbability.

\subsection{Is the hourglass an artifact of the tracer set?}

The objection. The Fields Medal is a narrow inclusion criterion applied at the present-day end. Since scholars can have several recorded mentors, lineages should fan out as they run backward from a small set of seeds, and contract again where records thin. An hourglass is what this shape looks like. The constriction may be at Leibniz for reasons of arithmetic rather than of history.

What we concede. The general shape is partly a consequence of the design. Backward traversal from 64 seeds through a multiparent DAG will broaden and then narrow, and one should not be impressed by broadening and narrowing as such.

What survives. The hourglass is not one shape; it is a family of shapes that differ node by node, and the differences are informative even if the general form of the family is not. Gauss, Hilbert, and Euler show thickness ratios of 15.1, 18.5, and 14.6, all constriction-and-expansion profiles. The pattern is a property of hub nodes in general, not of Leibniz specifically, and that is precisely the point: the tracer-set explanation predicts a global shape, but the variation between nodes is what carries information. Newton, on the same graph under the same traversal with the same seeds, shows no constriction at all. If the hourglass were purely a design artifact, Newton would have one too.

What distinguishes Leibniz within the family is not the ratio, where Hilbert is higher, but the combination of a high ratio with a large absolute path count: 47 of 64 lineages, against four for Newton. Ratio and volume are separately obtainable from the design; their conjunction is what we are pointing at.

What would settle it. The natural test is the one flagged in Section 5.2: a systematic comparison of thickness ratio and path volume across all high-traffic nodes in each century, so that the position of Leibniz within the distribution is visible rather than asserted. Our deposited data support this and we have not done it.

\subsection{Is the path concentration an artifact of the traversal?}

The objection. Section 4.4 conceded that the engine imposes its own preference for short, direct paths. A tool that generates path dependence, applied to a question about historical path dependence, is circular. This is the objection we take most seriously, and we answer it in four ways, in decreasing order of abstraction.

Scale. The preference of the engine operates at per-step deviations on the order of one part in a thousand. The historical concentration operates at 54 of 64 lineages, and at ratios of 10:1 and 46:1. These are not quantities a fourth-decimal-place per-step effect could produce.

Shape. The decay of the engine is smooth and monotonic in generational distance. The historical findings are discrete discontinuities: a wall at the eleventh century, a bottleneck at five twelfth-century individuals, a reorganization around 1700. The internal dynamics of the engine have no mechanism that would generate a discontinuity at a particular date.

Switchability. This is the decisive one. Rescue mode repairs every collision, pins CR1 to 1.0, and eliminates the parsimony penalty entirely, reproducing the hash-table baseline bitwise across all enumerated paths. Run in that mode, the traversal produces the same Leibniz hourglass, the same seven-indicator reorganization, the same upstream convergence on the five, and the same Monastery Wall. Every macro-finding in Section 5 is visible with the mechanism switched off.

Direction. The mechanism acts on the ordering of candidates within each frontier, not on the selection of the frontier itself. It refines what the traversal reports; it does not determine what the traversal reaches.

Why the mechanism is nonetheless worth having. Switching it off has a cost, and the cost is instructive. In Rescue mode every candidate has an identical cumulative score, so the ordering within the frontier falls back on the lexicographic order of identifiers, a criterion with no historical content. The consequence is visible in Tab. 12.

\medskip
\noindent \textbf{Tab. 12 Top-20 nodes by vote count, with and without the parsimony penalty.} One path traversal equals one vote. Both columns traverse identical data to identical depth. Entries in the right column marked with an occupation are the figures that displace the analytical mainstream under lexicographic ordering.

\begin{center}\small
\begin{tabular}{@{}p{0.084\linewidth}p{0.338\linewidth}p{0.084\linewidth}p{0.423\linewidth}@{}}
\toprule
\textbf{Rank} & \textbf{With penalty (N = 128, m = 27)} & \textbf{Rank} & \textbf{Lexicographic fallback} \\
\midrule
1 & J. Thomasius (d. 1684) & 1 & J. Thomasius (d. 1684) \\
2 & Leibniz (d. 1716) & 2 & B. Meisner (d. 1626) \\
3 & B. Meisner (d. 1626) & 3 & C. A. Hausen (d. 1743) \\
4 & W. of Ockham (d. 1349) & 3 & A. G. Kästner (d. 1800) \\
5 & Duns Scotus (d. 1308) & 5 & A. Rhode (d. 1633) \\
6 & A. Hegius (d. 1498) & 6 & A. Hegius (d. 1498) \\
7 & Euler (d. 1783) & 7 & W. of Ockham (d. 1349) \\
7 & J. Bernoulli (d. 1748) & 8 & Duns Scotus (d. 1308) \\
9 & T. à Kempis (d. 1471) & 9 & J. Dubois (d. 1555), surgery \\
10 & G. Groote (d. 1384) & 10 & T. à Kempis (d. 1471) \\
11 & Gonsalvus (d. 1313) & 11 & G. Groote (d. 1384) \\
12 & P. Olivi (d. 1298) & 12 & Gonsalvus (d. 1313) \\
13 & C. A. Hausen (d. 1743) & 13 & P. Olivi (d. 1298) \\
14 & A. G. Kästner (d. 1800) & 14 & J. Peckham (d. 1292) \\
15 & A. Rhode (d. 1633) & 15 & J. W. von Andernach (d. 1574), medicine \\
16 & J. Peckham (d. 1292) & 16 & H. Fabricius (d. 1619), anatomy \\
17 & Lagrange (d. 1813) & 17 & G. Falloppio (d. 1562), anatomy \\
18 & J. Martini (d. 1649) & 18 & Bonaventure (d. 1274) \\
19 & Bonaventure (d. 1274) & 19 & J. Lefèvre d'Étaples (d. 1536), humanities \\
20 & Poisson (d. 1840) & 20 & Leibniz (d. 1716) \\
\bottomrule
\end{tabular}\end{center}

Under lexicographic ordering the calculus lineage is replaced by anatomists, surgeons, physicians, and humanists whose vote counts happen to survive an alphabetical cut, and Leibniz is demoted to rank 20. Unbinding the FIELD component from the bundled representation of each top-20 set makes the difference measurable: the lexicographic set has a centroid on which anatomy ranks first and mathematics fourth, whereas the penalized set has mathematics first.

The displaced figures are not errors in the data. They are real and heavily traversed nodes, but they sit at generational depths of 21 to 26, whereas the analytical mainstream sits at 9 to 12. They are ancient junctions through which many paths happen to run, not participants in the transmission the tracer set is tracing. Distinguishing the two requires an ordering criterion that is sensitive to path structure, and lexicographic ordering by definition is not.

The general moral. On data at a scale exceeding the reach of case-study investigation, path dependence can arise not only from historical contingency but from the measurement architecture itself. The constructive response is not to hope the architecture is innocent, nor to suppress its distortions, but to characterize them in closed form and demonstrate what survives their removal. That is what Sections 4.4 and 6.3 have attempted.

\subsection{Is the predicate stable enough to carry this weight?}

The objection. Section 2 established that mentor and student name very different social relations in the eleventh, fifteenth, and twentieth centuries, and in European, Islamic, and Byzantine contexts. A single Wikidata predicate encodes all of them, and our method then treats them as comparable units of a path.

We do not have a full answer. This is the limitation we regard as most serious, and we do not want to minimize it. Two partial responses and one proposal follow.

The claims are calibrated to it. We have avoided any interpretation requiring generational distance to be a comparable quantity across periods. The findings that survive are ones about where the record concentrates, which is a claim about documentary structure, not about the intensity or nature of instruction.

The three predicates can be separated. P184, P802, and P1066 are not equivalent, and P184 in particular carries the modern doctoral meaning. We merged them for traversal because the coverage of any one alone is too sparse upstream. But the merge is reversible in the data: one could rerun the traversal on P184 alone and compare. We expect it to terminate far earlier, roughly at the nineteenth century, and that comparison would itself be informative about where the modern idiom stops applying.

A proposal. The right treatment is a period- and region-specific typology of the tie: doctoral supervision, \textit{licentia docendi}, \textit{ij\={a}za}, collegiate cohabitation, court patronage. Ties would be typed rather than merged, and paths would carry a record of which types they traverse. Constructing such a typology at 470,000-record scale is not something we can do by hand, and doing it automatically would require exactly the kind of institutional data that register-based studies possess and Wikidata does not.\footnote{De la Croix and Morault, ``Winners and Losers''; de la Croix, Scebba, and Zanardello, ``Flora, Cosmos, Salvatio.''} We suggest it as the most valuable point of collaboration between the two source traditions.

\section{Discussion}

\subsection{What the observations are for}

The three structural observations of Section 5 are not answers. They are, we hope, well-specified starting points, and it is worth saying explicitly what each opens.

The upstream convergence raises a question that has an active literature: what was the relation between non-Western scholarship and later Western scientific dominance? Our result is a topological fact about a mentorship record; it does not adjudicate that question. But it suggests a tractable next step that our data support and we have not taken. Track where the descendants of the five upstream hubs were located, institutionally and geographically, in each successive generation. If the center of gravity of that descent shifts, and if the shift can be dated, one would have a structural chronology of when scientific leadership moved, rather than only the observation that it did.\footnote{Maarten Bosker, Eltjo Buringh, and Jan Luiten van Zanden, ``From Baghdad to London: Unraveling Urban Development in Europe, the Middle East, and North Africa, 800--1800,'' Review of Economics and Statistics 95, no. 4 (2013): 1418--1437; Eric Chaney, ``Religion and the Rise and Fall of Islamic Science'' (working paper, Harvard University, May 2016; rev. March 2023); Timur Kuran, The Long Divergence: How Islamic Law Held Back the Middle East (Princeton: Princeton University Press, 2011).}

The seventeenth-century reorganization raises the question of institutional form. Our society-membership indicator is a count of P463 links, and it does not distinguish the kinds of institution that mattered most. Universities were local: they existed to teach, and their professors were expected to reside. Academies reached much further, and a substantial fraction of their membership corresponded from a distance, sometimes from another country. A lineage running through a university chair transmits differently from one running through the corresponding membership of an academy. The Leibniz--Newton contrast may bear on this directly, since Leibniz was an academician and founder of academies while Newton was an academy member who also held a Cambridge chair. Decomposing our membership surge by institution type is an exercise the deposited data support, and the institutional taxonomy it would require already exists.\footnote{Zanardello, ``Early Modern Academies''; de la Croix, Scebba, and Zanardello, ``Flora, Cosmos, Salvatio.''}

The Monastery Wall raises the question of what a change in recording medium looks like from inside the record, a question of general methodological interest to historical network research well beyond this corpus. The eleventh century is one such transition. Others, such as the professionalization of the doctorate around 1800 and the mid-twentieth-century expansion of graduate education, should leave analogous signatures, and testing whether they do is a way of validating the reading we propose in Section 5.3.

\subsection{Limitations}

Beyond those already stated in place, four. One channel: mentorship is one mode of transmission among several. The Republic of Letters moved ideas by correspondence; apprenticeship moved technical skill through a channel our framing sets aside entirely, and on which there is a substantial literature.\footnote{On the apprenticeship channel see the body of work by Patrick Wallis and coauthors on premodern skill transmission.} Institutional co-affiliation is a further channel. Our semantic recovery of the translation channel (Section 5.4) is indirect and weak. We have mapped one channel, not the network of transmission.

Institutional heterogeneity: as above, our indicators do not distinguish institution types that differed in reach and in mode of participation.

The cycles: we removed 369 bidirectional pairs as a precondition of traversal. We did not examine them case by case. Some are surely data errors; some may be genuine reciprocal or successive relations, which are historically possible. Inspecting them is a small, bounded task with real interpretive value, and we have not done it.

Validation: we have no independently verified sample against which to estimate link precision. This is the gap that most limits what our figures can be asked to bear.

\section{Conclusion}

Working from all 64 historical Fields Medalists as a fixed, ex ante tracer set on the integrated Wikidata mentor--student graph, we have enumerated a 25.5-million-path directed acyclic graph connecting modern research lineages to their premodern antecedents, and reported three structural observations: an hourglass concentration of documented traffic at Leibniz; a simultaneous reorganization of seven independently extracted predicate dimensions across a window centered on the same figure; and an upstream convergence of 84 percent of the traced lineages on five twelfth- and thirteenth-century Islamic and Byzantine scholars, terminating at an eleventh-century boundary.

The observations are about a graph as particular sources constitute it. Half of this paper is therefore given over to what that qualification costs: to the notability threshold that decides who appears, the crowdsourcing that decides what is asserted, the predicate that flattens a millennium of different social arrangements into one relation, and the traversal that decides which of 25 million paths a reader is shown.

Two things we would like carried out of the paper independently of its findings. The first is that a traversal at this scale must be audited, and that auditing is possible. The engine we used is algebraically reversible, which means that its own preference for short, direct paths could be quantified in closed form, switched off, and shown not to generate the macro-structures it refines. That demonstration is not incidental to the historical argument; without it, the argument would be unfalsifiable.

The second is that the natural output of such a traversal is a family of structures indexed by resolution, not a single ranked list. The same corpus, traversed by the same algebra at different resolutions, returns the Gauss lineage and the Leibniz--Euler lineage as mutually exclusive but individually coherent macro-structures. Reporting the family, with the parameter that selects among its members, is a strictly more informative act than reporting one member, and it is the kind of report that a method presenting a single plausible answer cannot make.

Where the record stops, and why it stops there, turned out to be as informative as what it contains. We take that to be a general property of network reconstruction from historical sources, and not a peculiarity of this one.

\section*{Data Accessibility Statement}

The underlying data are drawn from Wikidata and are publicly available under a CC0 license. The Wikidata extraction queries (SPARQL), the traversal configurations, the reference-vector specifications, the complete 64-member tracer set, the reference implementation of the traversal engine, and all tabular results reported here will be deposited in a public repository with a DOI on publication, and are available from the corresponding author on request in the meantime. All results are produced by a deterministic algorithm with no random seeds, so the reported figures are exactly reproducible from the deposited code and inputs. We note one caveat that follows from Section 4.5: results are configuration-relative, and a reader rerunning the traversal at a different resolution should expect a different within-frontier ranking. The provenance table in Section 4.6 is intended to make every reported figure rederivable.

\section*{Competing Interests}

The traversal architecture used in this study (VaCoAl) has been commercialized in hardware form by Shuhari System under the product name CASRAM. Yoichi Sato is affiliated with Shuhari System; Kanji Otsuka and Hiroyuki Chuma are among the originators of the architecture, and all three authors may benefit from its commercial development. The historical analysis reported here uses the software implementation only; all results are reproducible from the deposited code and inputs, and Section 6.3 reports a comparison against a standard hash-table baseline that a reader can run independently. The authors declare no other competing interests.

\section*{Authors' Contributions}

Hiroyuki Chuma: conceptualization, methodology, software, formal analysis, investigation, data curation, writing of the original draft, and writing, review, and editing. Kanji Otsuka: methodology of the traversal architecture, and writing, review, and editing. Yoichi Sato: methodology of the traversal architecture, and writing, review, and editing.

\section*{Acknowledgements}

We thank the Wikidata, Mathematics Genealogy Project, and MacTutor History of Mathematics Archive communities for maintaining the data infrastructure on which this study rests. We used a large language model for Japanese-to-English translation and editorial assistance, and AI-assisted development environments in building the traversal application. The authors retain full responsibility for all claims presented here. This research received no specific grant from any funding agency in the public, commercial, or not-for-profit sectors.

\section*{Appendix A. Extraction and Construction Detail}

From the Wikidata SPARQL endpoint (snapshot accessed 2026) we extracted all entities linked by doctoral advisor (P184), student (P802), and student of (P1066) within the component reachable from mathematics-oriented persons. Approximately 470,000 assertions resulted. A pre-purification pass removed 369 bidirectional pairs, yielding a directed acyclic graph of 372,853 persons.

For the 1,043 hub nodes identified from path traffic, up to twelve further predicate dimensions were attached by SPARQL query: field of work (P101), languages used (P1412), employer (P108), member of (P463), place of birth (P19), birth and death dates (P569, P570), and others. These were encoded as high-dimensional binary vectors and combined by binding and bundling (Appendix B). The complete traversal, including the semantic stage, completed in 20 to 30 minutes on a 2014-vintage Intel Xeon E5-1650 v3 with no GPU.

\section*{Appendix B. The Traversal Engine, in Brief}

This appendix gives what a reader needs in order to assess Sections 4 and 6.3. The full specification is given elsewhere.\footnote{Full specification in Chuma, Otsuka, and Sato, arXiv:2604.11665.}

Representation. Each person and each attribute value is a binary vector of approximately one million dimensions. Two operations combine them: binding, implemented as XOR with a cyclic shift over GF(2), which attaches a value to a role; and bundling, implemented as componentwise majority, which superposes several bound pairs into one composite. Unbinding is the inverse of binding and is exact, because XOR is its own inverse. Similarity is Hamming distance; the expected similarity between uncorrelated vectors is 0.5, which is the baseline against which every semantic measurement in this paper is reported.

Addressing. Rather than the probabilistic random projections of the standard approach, addresses are generated by deterministic diffusion over a Galois field: a polynomial remainder operation implementable as a linear-feedback shift register. The point of the finite field is that operations cycle within a bounded address space, and it is this cycling that scatters similar inputs into mutually unrelated directions, producing quasi-orthogonality deterministically rather than by sampling. A single-bit input change flips approximately half the output bits.

Block structure and Don't Care. Each representation is partitioned across N independent blocks, 64 to 1,024 in our sweeps, each of which maps its segment to an address and votes for the entry stored there. The winner is by majority. When two distinct items map to the same block address, that block is marked Don't Care and its vote is withheld rather than being allowed to produce a false majority.

The two modes. In Rescue mode, a four-stage pipeline intercepts and resolves every collision, pinning the per-step agreement rate CR1 to 1.0 and producing output bitwise identical to a hash-table baseline, verified across all enumerated paths. In Don't Care mode the pipeline is disabled, residual collisions are tolerated, and the cumulative product CR2 becomes the criterion by which within-frontier candidates are ranked. Section 6.3 reports the comparison; the main analysis uses Don't Care mode.

The bias. CR2(n) = CR2(n$-$1) $\times$ CR1(n$-$1). With CR1 approximately 0.997, the closed-form prediction at n = 56 is 0.846, while the measured value is 0.905, the gap being consistent with survivorship along deep paths (Fig. 1). Across the resolution sweep the closed-form endpoints range from 0.91 at N = 64 to 0.21 at N = 1024 (Fig. 2 and Tab. 3). The practical effect in every case is a mild preference for short, direct routes over long, circuitous ones.

\section*{Appendix C. Robustness}

Frontier size. Results are qualitatively stable across FS values of 10,000, 15,000, and 20,000. At FS = 25,000 the hash-table baseline fails with an out-of-memory error at 128 GB while the engine completes. The upstream convergence figure rises from approximately 72 percent at FS = 5,000 to 84 percent at FS of 10,000 or more, and plateaus thereafter.

Mode. Between Don't Care and Rescue modes, all macro-findings are preserved: the 84 percent convergence, the 10:1 hourglass ratio, the Monastery Wall, the 46-fold membership rise, and the composite rank of 91 for Newton. The difference appears in the within-frontier ranking of subthreshold candidates, precisely where CR2 carries content (Tab. 12).

Resolution. Reported in Section 4.5. Reachable depth is nonmonotonic in N: 55 generations at N = 64, 128, and 256; 57 at N = 512; and 56 at N = 1,024. N = 512 is the only configuration matching the maximum depth attained by the hash-table baseline at FS = 20,000.

Tracer set. A full replication with an alternative tracer set, Nobel laureates in physics for instance, is beyond the present scope. The structural logic, that a fixed ex ante tracer set induces a well-defined subgraph on a DAG too large for unaided inspection, does not depend on the choice of the Fields Medal, and the predicates, the DAG structure, and the FS relativity apply identically. A different tracer set would yield different chokepoint identities. We regard establishing whether it yields a differently shaped concentration as the most informative single replication available.

\section*{Bibliography}
\begingroup\setlength{\parindent}{-1.5em}\setlength{\leftskip}{1.5em}\small

Almelhem, Ali, Murat Iyigun, Austin Kennedy, and Jared Rubin. ``Enlightenment Ideals and Belief in Progress in the Run-up to the Industrial Revolution: A Textual Analysis.'' Quarterly Journal of Economics 141, no. 1 (2026): 263--314. \url{https://doi.org/10.1093/qje/qjaf054.}\par

Antognazza, Maria Rosa. Leibniz: A Very Short Introduction. Oxford: Oxford University Press, 2016.\par

Bosker, Maarten, Eltjo Buringh, and Jan Luiten van Zanden. ``From Baghdad to London: Unraveling Urban Development in Europe, the Middle East, and North Africa, 800--1800.'' Review of Economics and Statistics 95, no. 4 (2013): 1418--1437. \url{https://doi.org/10.1162/REST_a_00284.}\par

Cabello, Matias. ``The Counter-Reformation, Science, and Long-Term Growth: A Black Legend?'' SSRN Working Paper 4389708, 2023. \url{https://doi.org/10.2139/ssrn.4389708.}\par

Chaney, Eric. ``Religion and the Rise and Fall of Islamic Science.'' Working paper, Harvard University, May 2016; revised March 2023.\par

de la Croix, David, and Marc Goñi. ``Nepotism vs. Intergenerational Transmission of Human Capital in Academia (1088--1800).'' Journal of Economic Growth 29, no. 4 (2024): 469--514. \url{https://doi.org/10.1007/s10887-024-09244-0.}\par

de la Croix, David, and Pauline Morault. ``Winners and Losers from the Protestant Reformation: An Analysis of the Network of European Universities.'' Journal of Economic History 86, no. 2 (2026): 460--500. \url{https://doi.org/10.1017/S0022050725101058.}\par

de la Croix, David, Rossana Scebba, and Chiara Zanardello. ``Flora, Cosmos, Salvatio: Pre-modern Academic Institutions and the Spread of Ideas.'' CEPR Discussion Paper DP20569, 2025.\par

de Ridder-Symoens, Hilde, ed. A History of the University in Europe. Vol. 1, Universities in the Middle Ages. Cambridge: Cambridge University Press, 1992.\par

Düring, Marten, and Linda von Keyserlingk. ``Netzwerkanalyse in den Geschichtswissenschaften: Historische Netzwerkanalyse als Methode für die Erforschung von historischen Prozessen.'' In Prozesse: Formen, Dynamiken, Erklärungen, edited by Rainer Schützeichel and Stefan Jordan, 337--350. Wiesbaden: Springer VS, 2015. \url{https://doi.org/10.1007/978-3-531-93458-7_15.} \url{https://doi.org/10.1007/978-3-531-93458-7_15.}\par

Grajzl, Peter, and Peter Murrell. ``A Macroscope of English Print Culture, 1530--1700, Applied to the Coevolution of Ideas on Religion, Science, and Institutions.'' Social Science History 48 (2024): 489--519. \url{https://doi.org/10.1017/ssh.2024.17.}\par

Hall, A. Rupert. Philosophers at War: The Quarrel between Newton and Leibniz. Cambridge: Cambridge University Press, 1980.\par

Itoh, Shuntaro. The Twelfth-Century Renaissance: The World of Arabic Civilisation and Western Europe. Tokyo: Iwanami Shoten, 1993. [In Japanese.]\par

Jaeger, C. Stephen. The Envy of Angels: Cathedral Schools and Social Ideals in Medieval Europe, 950--1200. Philadelphia: University of Pennsylvania Press, 2000.\par

Koschnick, Julius. ``Teacher-Directed Scientific Change: The Case of the English Scientific Revolution.'' EHES Working Paper 274, 2025.\par

Kuran, Timur. The Long Divergence: How Islamic Law Held Back the Middle East. Princeton: Princeton University Press, 2011.\par

Laouenan, Morgane, Palaash Bhargava, Jean-Benoît Eyméoud, Olivier Gergaud, Guillaume Plique, and Etienne Wasmer. ``A Cross-Verified Database of Notable People, 3500BC--2018AD.'' Scientific Data 9, no. 1 (2022): 290. \url{https://doi.org/10.1038/s41597-022-01369-4.}\par

Lemercier, Claire. ``Formal Network Methods in History: Why and How?'' In Social Networks, Political Institutions, and Rural Societies, edited by Georg Fertig, 281--310. Turnhout: Brepols, 2015. \url{https://doi.org/10.1484/M.RURHE-EB.4.00198.} \url{https://doi.org/10.1484/M.RURHE-EB.4.00198.}\par

Mokyr, Joel. A Culture of Growth: The Origins of the Modern Economy. Princeton: Princeton University Press, 2016.\par

Myers, Sean A., Peter J. Mucha, and Mason A. Porter. ``Mathematical Genealogy and Department Prestige.'' Chaos 21, no. 4 (2011): 041104. \url{https://doi.org/10.1063/1.3668043.}\par

Rashdall, Hastings. The Universities of Europe in the Middle Ages. 3 vols. 1895. Reprint, Cambridge: Cambridge University Press, 2010.\par

Rollinger, Christian. ``Prolegomena: Problems and Perspectives of Historical Network Research and Ancient History.'' Journal of Historical Network Research 4 (2020): 1--35.\par

Saliba, George. Islamic Science and the Making of the European Renaissance. Cambridge [MA]: MIT Press, 2007.\par

Serafinelli, Michel, and Guido Tabellini. ``Creativity over Time and Space: A Historical Analysis of European Cities.'' Journal of Economic Growth 27, no. 1 (2022): 1--43. \url{https://doi.org/10.1007/s10887-021-09199-6.}\par

Squicciarini, Mara P., and Nico Voigtländer. ``Human Capital and Industrialization: Evidence from the Age of Enlightenment.'' Quarterly Journal of Economics 130, no. 4 (2015): 1825--1883.\par

Stelter, Robert, and Diego Alburez-Gutiérrez. ``Representativeness Is Crucial for Inferring Demographic Processes from Online Genealogies: Evidence from Lifespan Dynamics.'' Proceedings of the National Academy of Sciences 119, no. 10 (2022): e2120455119. \url{https://doi.org/10.1073/pnas.2120455119.}\par

Weber, Max. ``Science as a Vocation.'' 1919. In The Vocation Lectures, edited by David Owen and Tracy B. Strong. Indianapolis: Hackett, 2004.\par

Zanardello, Chiara. ``Early Modern Academies, Universities, and Economic Growth.'' LIDAM Discussion Paper 2024/12, 2024.\par

\endgroup
\end{document}